\documentclass[prb,aps,twocolumn,showpacs,superscriptaddress]{revtex4}
\usepackage{graphicx,amsmath,bm}
\usepackage{dcolumn}

\begin{document}

\title{Exact results on the two-particle Green's function of a Bose-Einstein condensate}

\author{Takafumi Kita}
%\email{}
\affiliation{Department of Physics, Hokkaido University,
Sapporo 060-0810, Japan}
\date{\today}

\begin{abstract}
Starting from the Dyson-Beliaev and generalized Gross-Pitaevskii equations with an extra nonlocal potential, we derive an exact expression of the two-particle Green's function $\underline{\cal K}$ for an interacting Bose-Einstein condensate in terms of unambiguously defined self-energies and vertices. The formula can be a convenient basis for approximate calculations of  $\underline{\cal K}$. It also tells us that poles of $\underline{\cal K}$ are not shared with (i.e.\ shifted from) those of the single-particle Green's function, contrary to the conclusion of previous studies.
\end{abstract}


\maketitle

The realization \cite{AEMWC95} of Bose-Einstein condensation (BEC) with an atomic gas  in 1995 has revived
intense theoretical interests on interacting condensed Bose systems. 
One of their unique features is that 
the gapless Nambu-Goldstone boson \cite{PS95} of the broken U(1) symmetry, i.e.\ the Bogoliubov mode,\cite{Bogoliubov47} emerges as a pole of the single-particle Green's function $\hat{G}$ to dominate thermodynamic properties.
It also seems to have been widely accepted that poles of $\hat{G}$ are shared with those of the two-particle Green's function $\underline{\cal K}$, as first claimed by Gavoret and Nozi\`eres \cite{GN64}  in 1964 and reproduced by the dielectric formalism.\cite{SK74,WG74,Griffin93}
These theories have provided a support to utilize $\hat{G}$ for describing collective modes of condensed atomic gases.
Indeed, the sharing of common poles between $\hat{G}$ and $\underline{\cal K}$ has been regarded as one of the most spectacular features of condensed Bose systems.

However, the theory by Gavoret and Nozi\`eres \cite{GN64} is based on an analysis of the structures of simple perturbation expansions performed separately for $\hat{G}$ and $\underline{\cal K}$. Thus, it may suffer from ambiguity as to how to define self-energies and vertices in the presence of an ``improper'' interaction having only a single quasiparticle channel inherent in BEC. Since BEC is a prototype of broken symmetry, it will be well worth reinvestigating the fundamental issue with a different method and viewpoint. 

As is well known in normal systems,\cite{MS59,BK61,Baym62} a two-particle Green's function can be generated from a single-particle Green's function by a functional differentiation with respect to an additional potential. This method enables us to derive a formally exact expression of the two-particle Green's function in terms of unambiguously defined self-energies and vertices. Moreover, it can be used in practical calculations of the two-particle Green's function with Baym's $\Phi$-derivable approximation.\cite{Baym62} The approximation has a great advantage that the whole series of thermodynamic, single-particle, and two-particle properties  can be discussed in a unified way based on a single functional $\Phi$, even beyond equilibrium.\cite{Kita10}

We here apply the functional-differentiation method to an interacting Bose-Einstein condensate to obtain an exact expression of $\underline{\cal K}$. The formula can also be used for practical calculations of $\underline{\cal K}$ with the self-consistent $\Phi$-derivative approximation of condensed Bose systems developed recently.\cite{Kita09} Our derivation is based solely on rigorous results of the previous paper.\cite{Kita09} It will thereby be shown that poles of $\underline{\cal K}$ are not shared with those of $\hat{G}$, contrary to the previous conclusion. \cite{GN64,SK74,WG74,Griffin93} Unlike the previous studies for homogeneous systems using the momentum conservation,\cite{GN64,SK74,WG74,Griffin93} our formulation will be carried out in the coordinate space so that it is applicable to trapped atomic gases.

We consider identical Bose particles with mass $m$ and spin $0$
described by the Hamiltonian:
\begin{eqnarray}
H&=&\int d^{3}r_{1} \psi^{\dagger}({\bf r}_{1})K_{1}\psi({\bf r}_{1})
+\frac{1}{2}\int d^{3}r_{1}\int d^{3}r_{2}\,\psi^{\dagger}({\bf r}_{1})
\nonumber \\
& & \times \psi^{\dagger}({\bf r}_{2})V({\bf r}_{1}-{\bf r}_{2})
 \psi({\bf r}_{2})\psi({\bf r}_{1}) .
\label{H}
\end{eqnarray}
Here $\psi^{\dagger}$ and $\psi$ are field operators satisfying the Bose commutation relations, 
$K_{1}\equiv -{\hbar^{2}\nabla_{1}^{2}}/{2m}-\mu$
with $\mu$ the chemical potential,
and $V$ is the interaction potential. 
Though dropped here, the effect of a trap potential can be included easily in $K_{1}$.
Let us introduce the Heisenberg representations of the field operators by
\begin{equation}
\psi_{1}(1) \equiv e^{\tau_{1}H}\psi({\bf r}_{1})e^{-\tau_{1}H},\hspace{3mm}
\psi_{2}(1) \equiv e^{\tau_{1}H}\psi^{\dagger}({\bf r}_{1})e^{-\tau_{1}H},
\end{equation}
with $1\equiv ({\bf r}_{1},\tau_{1})$, where $0 \leq \tau_{1} \leq T^{-1}$ with 
$T$ the temperature in units of $\hbar=k_{\rm B}=1$.
The operators $\psi_{1}(1)$ and $\psi_{2}(1)$ were denoted previously\cite{Kita09} by $\psi(1)$ and $\bar{\psi}(1)$,
respectively.
We next express $\psi_{i}(1)$
as a sum of the condensate wave function $\Psi_{i}(1)\equiv \langle
\psi_{i}(1)\rangle$ and the quasiparticle field $\phi_{i}(1)$
as 
\begin{equation}
\psi_{i}(1)=\Psi_{i}(1)+\phi_{i}(1)\hspace{5mm}(i=1,2),
\label{psi=Psi+psi-t}
\end{equation}
with $\langle\cdots\rangle$ the grand-canonical average in terms of $H$. 
Note: (i) $\langle \phi_{i}(1)\rangle=0$ by definition; and (ii)  $\Psi_{1}(1)=\Psi_{2}^{*}(1)=\Psi({\bf r}_{1})$ in equilibrium with the superscript $^{*}$ signifying complex conjugate.
Using $\phi_{i}$, we introduce our Matsubara 
Green's function in the $2\times 2$ Nambu space by \cite{Kita09}
\begin{equation}
G_{ij}(1,2)\equiv - \bigl< T_{\tau}
\phi_{i}(1)\phi_{3-j}(2)\bigr> (-1)^{j-1} ,
\label{hatG}
\end{equation}
where $T_{\tau}$ denotes the ``time''-ordering operator. \cite{AGD63}
They satisfy \cite{Kita09}
\begin{eqnarray}
G_{ij}(1,2)\!\! &=&\!\! (-1)^{i+j-1}G_{3-j,3-i}(2,1)
\nonumber \\
\!\! &=&\!\! (-1)^{i+j}G_{ji}^{*}({\bf r}_{2}\tau_{1},{\bf r}_{1}\tau_{2}).
\label{G_ij-symm}
\end{eqnarray}

Let us recapitulate exact results on the matrix 
$\hat{G}=(G_{ij})$ and the vector $\vec{\Psi}=[\Psi_{1}\,\Psi_{2}]^{\rm T}$; see Sec.\ II of Ref.\ \onlinecite{Kita09} for details.
First of all, they obey
the Dyson-Beliaev equation and 
the generalized Gross-Pitaevskii equation (or generalized Hugenholtz-Pines relation)
given by
\begin{subequations}
\label{Dyson1}
\begin{equation}
\hat{G}^{-1}(1,\bar{3})\hat{G}(\bar{3},2)
=\hat{\sigma}_{0}\delta(1,2) ,
\end{equation}
\begin{equation}
\hat{G}^{-1}(1,\bar{2})\hat{\sigma}_{3}\vec{\Psi}(\bar{2})  =\vec{0} ,
\label{GP}
\end{equation}
\end{subequations}
respectively.
Here summations over barred arguments are implied, $\hat{\sigma}_{0}$ and $\hat{\sigma}_{3}$ denote the $2\times 2$ unit matrix and the third Pauli matrix, respectively, 
$\delta(1,2)\equiv \delta(\tau_{1}-\tau_{2})\delta({\bf r}_{1}-{\bf r}_{2})$,
and $\hat{G}^{-1}$ is defined by
\begin{equation}
\hat{G}^{-1}(1,2)\equiv \left(-\hat{\sigma}_{0}\frac{\partial}{\partial \tau_{1}}
-\hat{\sigma}_{3}K_{1}\right)\delta(1,2)
-\hat{\Sigma}(1,2) ,
\label{G^-1(1,2)}
\end{equation}
with $\hat{\Sigma}$ the self-energy matrix. We point out that the first component of Eq.\ (\ref{GP}) in equilibrium is written explicitly as 
$-K_{1}\Psi({\bf r}_{1})=\Sigma_{11}(1,\bar{2})\Psi(\bar{\bf r}_{2})-\Sigma_{12}(1,\bar{2})\Psi^{*}(\bar{\bf r}_{2})$.
By approximating $\Sigma_{11}(1,2)= 2g\delta(1,2)|\Psi({\bf r}_{1})|^{2}$ and $\Sigma_{12}(1,2)= g\delta(1,2)[\Psi({\bf r}_{1})]^{2}$ for $V({\bf r}_{1}-{\bf r}_{2})=g\delta({\bf r}_{1}-{\bf r}_{2})$, it reduces to the standard Gross-Pitaevskii equation.\cite{Gross61,Pitaevskii61,PS08}
Setting $\Psi\rightarrow\sqrt{n}_{0}$ and $K\rightarrow -\mu$ with $n_{0}$ the condensate density in the same equation, we also obtain the Hugenholtz-Pines relation for the homogeneous system.\cite{HP59,Kita09}

It has been shown\cite{Kita09} that the elements of $\hat{\Sigma}$ satisfy the same relations as Eq.\ (\ref{G_ij-symm}).
Moreover, all of them can be obtained from a single functional
$\Phi=\Phi[G,F,\bar{F},\Psi_{1},\Psi_{2}]$ as Eq.\ (21a) of Ref.\ \onlinecite{Kita09}
with $G=G_{11}$, $F=G_{12}$, and $\bar{F}=-G_{21}$.
Using Eq.\ (\ref{G_ij-symm}), we here write every $G$ in $\Phi$ as 
$G(1,2)=[G_{11}(1,2)-G_{22}(2,1)]/2$.
Then the relevant relations can be put into the single expression:
\begin{subequations}
\label{Sigma-Phi}
\begin{equation}
\Sigma_{ij}(1,2)=-\frac{2}{T}\frac{\delta\Phi}{\delta G_{ji}(2,1)}.
\label{Sigma-Phi1}
\end{equation}
The functional $\Phi$ also satisfies Eq.\ (21b) of Ref.\ \onlinecite{Kita09}, i.e.,
\begin{equation}
\frac{1}{T}\frac{\delta \Phi}{\delta \Psi_{3-i}(1)}= \Sigma_{i\bar{j}}(1,\bar{2})(-1)^{\bar{j}-1}\Psi_{\bar{j}}(\bar{2}) .
\label{Sigma-Phi2}
\end{equation}
\end{subequations}

With these preliminaries, we now study the two-particle Green's function:
\begin{eqnarray}
&&\hspace{-2mm}{\cal K}_{ij,kl}(12,34)
\nonumber \\
&&\hspace{-6mm}
\equiv \langle T_{\tau}\psi_{i}(1)\psi_{k}(3)\psi_{3-l}(4)\psi_{3-j}(2)\rangle (-1)^{j+l}
\nonumber \\
&&\hspace{-2mm}
-\langle T_{\tau}\psi_{i}(1)\psi_{3-j}(2)\rangle\langle T_{\tau}\psi_{k}(3)\psi_{3-l}(4)\rangle(-1)^{j+l} .
\label{calK}
\end{eqnarray}
Collective modes correspond to the poles of this Green's function.
To derive the equation for ${\cal K}$, 
we follow a standard procedure to produce the two-particle Green's function 
from $\hat{G}$. \cite{MS59,BK61}
Let us add an extra perturbation described by the $S$ matrix:
\begin{equation}
{\cal S}(\beta)\equiv T_{\tau}
\exp\!\left[-\frac{1}{2} \psi_{\,\bar{i}}(\bar{1})\psi_{3-\bar{j}}(\bar{2})(-1)^{\bar{j}-1} U_{\bar{j}\bar{i}}(\bar{2},\bar{1})\right]\! ,
\label{calS}
\end{equation}
with $\beta\equiv T^{-1}$.
The full Matsubara Green's function in the presence of the nonlocal potential $\hat{U}\equiv (U_{ij})$ is defined by \cite{AGD63}
\begin{subequations}
\begin{eqnarray}
{\cal G}_{ij}(1,2)\!\! &\equiv&\!\!  -\frac{\langle T_{\tau}{\cal S}(\beta)\psi_{i}(1)\psi_{3-j}(2)\rangle}
{\langle{\cal S}(\beta)\rangle} (-1)^{j-1}
\nonumber \\
\!\! &=&\!\! -\langle T_{\tau}{\cal S}(\beta)\psi_{i}(1)\psi_{3-j}(2)\rangle_{\rm c}(-1)^{j-1},
\label{calG-def1}
\end{eqnarray}
where the subscript $_{\rm c}$ denotes contribution of those Feynman diagrams 
connected with $\psi_{i}(1)$ and/or $\psi_{3-j}(2)$.
Noting that there may be the finite average
$\Psi_{i}(1)\equiv \langle T_{\tau} {\cal S}(\beta)\psi_{i}(1)\rangle_{\rm c}$,
we can transform Eq.\ (\ref{calG-def1}) into
\begin{equation}
{\cal G}_{ij}(1,2)=G_{ij}(1,2)-\Psi_{i}(1)\Psi_{3-j}(2)(-1)^{j-1},
\label{calG-def2}
\end{equation}
\end{subequations}
with $G_{ij}(1,2)\equiv -\langle T_{\tau} {\cal S}(\beta)\phi_{i}(1)\phi_{3-j}(2)\rangle_{\rm c}(-1)^{j-1}$; this
quantity reduces to Eq.\ (\ref{hatG}) as $\hat{U}\rightarrow\hat{0}$.
It may be seen easily that two-particle Green's function (\ref{calK}) is obtained from Eq.\ (\ref{calG-def1}) by
\begin{subequations}
\label{calK-deriv}
\begin{equation}
{\cal K}_{ij,kl}(12,34)=2\frac{\delta {\cal G}_{ij}(1,2)}{\delta U_{lk}(4,3)},
\label{calK-deriv1}
\end{equation}
where the limit $\hat{U}\rightarrow\hat{0}$ is implied after the differentiation; we will use this convention below.
A substitution of Eq.\ (\ref{calG-def2}) into Eq.\ (\ref{calK-deriv1}) yields
\begin{eqnarray}
&&\hspace{-5mm}
{\cal K}_{ij,kl}(12,34) 
= 2 \frac{\delta G_{ij}(1,2)}{\delta U_{lk}(4,3)}
-2\biggl[\Psi_{i}(1)\frac{\delta\Psi_{3-j}(2)}{\delta U_{lk}(4,3)}
\nonumber \\
&&\hspace{18.5mm}
+\frac{\delta \Psi_{i}(1)}{\delta U_{lk}(4,3)} \Psi_{3-j}(2)\biggr](-1)^{j-1} .
\label{calK-deriv2}
\end{eqnarray}
\end{subequations}
Equation (\ref{calK-deriv2}) tells us that we only need to know the linear responses of
$\hat{G}$ and $\vec{\Psi}$ to $\hat{U}$ for writing ${\cal K}$ down explicitly.

To carry it out, we start from Eq.\ (\ref{Dyson1}). Differentiations of $G_{ij}(1,2)\equiv -\langle T_{\tau} {\cal S}(\beta)\phi_{i}(1)\phi_{3-j}(2)\rangle_{\rm c}(-1)^{j-1}$ and $\Psi_{i}(1)\equiv \langle T_{\tau} {\cal S}(\beta)\psi_{i}(1)\rangle_{\rm c}$ with respect to $\tau_{1}$ tell us \cite{MS59,BK61} that
perturbation, Eq.\  (\ref{calS}), adds to the right-hand side of Eq.\ (\ref{G^-1(1,2)}) an extra term $-\hat{U}'(1,2)$ with
\begin{equation}
U_{ij}'(1,2)\equiv \frac{U_{ij}(1,2)+(-1)^{i+j-1}U_{3-j,3-i}(2,1)}{2} .
\label{U'}
\end{equation}
Varying $\hat{U}\rightarrow \hat{U}+\delta \hat{U}$ and subsequently setting $\hat{U}=\hat{0}$
in resultant Eq.\ (\ref{Dyson1}), we obtain the first-order equations,
\begin{subequations}
\label{Dyson2}
\begin{equation}
\hat{G}^{-1}(1,\bar{3})\delta\hat{G}(\bar{3},2)
=\bigl[\delta \hat{U}'(1,\bar{3}) +\delta \hat{\Sigma}(1,\bar{3})\bigr] \hat{G}(\bar{3},2),
\label{Dyson2a}
\end{equation}
\begin{equation}
\hat{G}^{-1}(1,\bar{2})\hat{\sigma}_{3}\delta\vec{\Psi}(\bar{2})
=\bigl[\delta \hat{U}'(1,\bar{2})+\delta \hat{\Sigma}(1,\bar{2})\bigr]\hat{\sigma}_{3}\vec{\Psi}(\bar{2}).
\label{Dyson2b}
\end{equation}
\end{subequations}

At this stage, it is convenient to introduce the following quantities:
\begin{subequations}
\label{Gamma}
\begin{equation}
\Gamma^{(4)}_{ij,kl}(12,34)\equiv -\frac{1}{2}\frac{\delta\Sigma_{ij}(1,2)}{\delta G_{lk}(4,3)}
=\frac{1}{T}\frac{\delta^{2}\Phi }{\delta G_{ji}(2,1)\delta G_{lk}(4,3)},
\label{Gamma(0)}
\end{equation}
\begin{eqnarray}
\Gamma^{(3)}_{ij,k}(12,3)\!\!&\equiv&\!\! \frac{1}{2}(-1)^{k-1}\frac{\delta\Sigma_{ij}(1,2)}{\delta \Psi_{k}(3)}
\nonumber \\
\hspace{4.7mm}
\!\!&=&\!\!2(-1)^{k+\bar{l}} \Gamma^{(4)}_{ij,3-k,\bar{l}}(12,3\bar{4})\Psi_{\bar{l}}(\bar{4}) ,
\label{Gamma(1)}
\end{eqnarray}
\begin{eqnarray}
\tilde{\Gamma}^{(3)}_{i,jk}(1,23)\!\!&\equiv&\!\! 2(-1)^{\bar{l}-1} \Gamma^{(4)}_{i\bar{l},jk}(1\bar{4},23)\Psi_{\bar{l}}(\bar{4}) 
\nonumber \\
\!\!&=&\!\!(-1)^{i}\Gamma^{(3)}_{jk,3-i}(23,1),
\label{Gamma(1)t}
\end{eqnarray}
\begin{equation}
\Gamma^{(2)}_{ij}(1,2)\equiv 2(-1)^{\bar{k}-1}\Gamma^{(3)}_{i\bar{k},j}(1\bar{3},2)\Psi_{\bar{k}}(\bar{3}),
\label{Gamma(2)}
\end{equation}
\end{subequations}
where Eq.\ (\ref{Sigma-Phi}) has been used to derive the second expression of $\Gamma^{(4,3)}$.
These are  ``irreducible'' vertices of our condensed Bose system, as seen below, and can be expressed diagrammatically as Fig.\ \ref{fig:Fig1}.
It follows from Eq.\ (\ref{G_ij-symm}) and $\Phi^{*}=\Phi$ that they satisfy various symmetry relations, e.g., 
$\Gamma^{(4)}_{ij,kl}(12,34)=\Gamma^{(4)}_{kl,ij}(34,12)=(-1)^{i+j-1}\Gamma^{(4)}_{3-j,3-i,kl}(21,34)$.
The quantities $\Gamma^{(4)}$ and $\Gamma^{(3)}$ correspond to $I$ and $J$ of Gavoret and Nozi\`eres, \cite{GN64} respectively.
Our definitions may be advantageous over theirs because the vertices can be obtained explicitly from a single functional $\Phi$
with clear relations among them.
\begin{figure}[t]
        \begin{center}
                \includegraphics[width=0.9\linewidth]{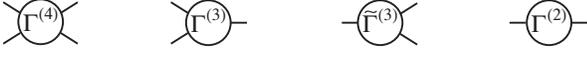}
        \end{center}
        \caption{Irreducible vertices of Eq.\ (\ref{Gamma}).}
        \label{fig:Fig1}
\end{figure}

Using $\Gamma^{(4,3)}$ above, we can express $\delta \hat{\Sigma}$ in Eq.\ (\ref{Dyson2}) as
\begin{eqnarray}
\delta \Sigma_{ij}(1,2)\!\! &=&\!\! -2\Gamma^{(4)}_{ij,\bar{l}\bar{k}}(12,\bar{4}\bar{3})\delta G_{\bar{k}\bar{l}}(\bar{3},\bar{4})
\nonumber \\
\!\! & &\!\! 
+2\Gamma^{(3)}_{ij,\bar{k}}(12,\bar{3})(-1)^{\bar{k}-1}\delta \Psi_{\bar{k}}(\bar{3}).
\label{dSigma}
\end{eqnarray}
It enables us to transform Eq.\ (\ref{Dyson2}) into a closed set of equations for 
$\delta\hat{G}$ and $\delta\vec{\Psi}$. Indeed, multiplying Eq.\ (\ref{Dyson2}) by $\hat{G}$ from the left,
substituting Eq.\ (\ref{dSigma}), and using Eqs.\ (\ref{Gamma(1)t}) and (\ref{Gamma(2)}), we obtain
\begin{subequations}
\label{dGPsi-eq}
\begin{eqnarray}
& &\!\!  \delta G_{ij}(1,2)
\nonumber \\
\!\!&=&\!\! G_{i\bar{l}}(1,\bar{4})G_{\bar{k}j}(\bar{3},2)\delta U_{\bar{l}\bar{k}}'(\bar{4},\bar{3})
\nonumber \\
& &\!\!-2G_{i\bar{l}}(1,\bar{4})
 G_{\bar{k}j}(\bar{3},2)\Gamma^{(4)}_{\bar{l}\bar{k},\bar{n}\bar{m}}(\bar{4}\bar{3},\bar{6}\bar{5})\delta G_{\bar{m}\bar{n}}(\bar{5},\bar{6})
\nonumber \\
& &\!\!+2G_{i\bar{l}}(1,\bar{4})G_{\bar{k}j}(\bar{3},2)\Gamma^{(3)}_{\bar{l}\bar{k},\bar{m}}(\bar{4}\bar{3},\bar{5})(-1)^{\bar{m}-1}  \delta \Psi_{\bar{m}}(\bar{5}),
\nonumber \\
\label{dG-eq}
\end{eqnarray}
\begin{eqnarray}
 (-1)^{i-1}\delta \Psi_{i}(1)
\!\!&=&\!\!
G_{i\bar{k}}(1,\bar{3})(-1)^{\bar{j}-1}\Psi_{\bar{j}}(\bar{2}) \delta U_{\bar{k}\bar{j}}'(\bar{3},\bar{2})
\nonumber \\
& &\!\!
-G_{i\bar{j}}(1,\bar{2})
\tilde{\Gamma}^{(3)}_{\bar{j},\bar{l}\bar{k}}(\bar{2},\bar{4}\bar{3})\delta G_{\bar{k}\bar{l}}(\bar{3},\bar{4})
\nonumber \\
& &\!\!
+G_{i\bar{j}}(1,\bar{2})\Gamma^{(2)}_{\bar{j}\bar{k}}(\bar{2},\bar{3})(-1)^{\bar{k}-1} \delta \Psi_{\bar{k}}(\bar{3}) .
\nonumber \\
\label{dPsi-eq}
\end{eqnarray}
\end{subequations}
Note $G_{i\bar{l}}(1,\bar{4})G_{\bar{k}j}(\bar{3},2)\delta U_{\bar{l}\bar{k}}'(\bar{4},\bar{3})
=\frac{1}{2}
[G_{i\bar{l}}(1,\bar{4})G_{\bar{k}j}(\bar{3},2)+(-1)^{\bar{k}+\bar{l}-1}G_{i,3-\bar{k}}(1,\bar{3})G_{3-\bar{l},j}(\bar{4},2)]
\delta U_{\bar{l}\bar{k}}(\bar{4},\bar{3})$ from Eq.\ (\ref{U'}). Using $\Gamma^{(4)}_{ij,kl}(12,34)=(-1)^{i+j-1}\Gamma^{(4)}_{3-j,3-i,kl}(21,34)$, we can also transform
 $G_{i\bar{l}}(1,\bar{4}) G_{\bar{k}j}(\bar{3},2)\Gamma^{(4)}_{\bar{l}\bar{k},mn}(\bar{4}\bar{3},56)
=(-1)^{\bar{k}+\bar{l}-1}G_{i,3-\bar{k}}(1,\bar{3})G_{3-\bar{l},j}(\bar{4},2)
\Gamma^{(4)}_{\bar{l}\bar{k},mn}(\bar{4}\bar{3},56)$.

To provide Eq.\ (\ref{dGPsi-eq}) with a compact expression, let us introduce the vectors $\delta\vec{G}$ 
and $\delta \vec{U}$ by
\begin{equation}
\langle 12_{ij}|\delta\vec{G}=\delta G_{ij}(1,2),
\hspace{5mm}
\langle 12_{ij}|\delta\vec{U}=\delta U_{ij}(1,2) ,
\label{Vec3}
\end{equation}
together with the matrices $\underline{\cal K}$, $\underline{\Gamma}^{(4)}$, 
$\underline{\chi}^{(0)}$, 
$\underline{1}$, $\underline{\Gamma}^{(3)}$, $\underline{\tilde{\Gamma}}^{(3)}$, 
$\underline{\Psi}^{(3)}$, $\underline{\tilde{\Psi}}^{(3)}$, and
$\hat{\Gamma}^{(2)}$ by
\begin{subequations}
\label{VecMat}
\begin{equation}
\langle 12_{ij}|\underline{\cal K}|43_{lk}\rangle\equiv
{\cal K}_{ij,kl}(12,34),
\label{calK-mat}
\end{equation}
\begin{equation}
\langle 12_{ij}|\underline{\Gamma}^{(4)}|43_{lk}\rangle\equiv
\Gamma^{(4)}_{ij,kl}(12,34),
\end{equation}
\begin{eqnarray}
\langle 12_{ij}|\underline{\chi}^{(0)}|43_{lk}\rangle\!\!&\equiv&\!\!
G_{il}(1,4)G_{kj}(3,2)+(-1)^{k+l-1}
\nonumber \\
& &\!\! \times G_{i,3-k}(1,3)G_{3-l,j}(4,2),
\label{chi^(0)}
\end{eqnarray}
\begin{equation}
\langle 12_{ij}|\underline{1}|43_{lk}\rangle\equiv 
\delta_{il}\delta_{kj}\delta(1,4)\delta(3,2),
\label{u1}
\end{equation}
\begin{equation}
\langle 12_{ij}|\underline{\Gamma}^{(3)}|3_{k}\rangle\equiv
\Gamma^{(3)}_{ij,k}(12,3),
\label{Gamma^(1)}
\end{equation}
\begin{equation}
\langle 1_{i}|\underline{\tilde{\Gamma}}^{(3)}|32_{kj}\rangle\equiv
\tilde{\Gamma}^{(3)}_{i,jk}(1,23),
\label{Gamma^(1)t}
\end{equation}
\begin{eqnarray}
\langle 12_{ij}|\underline{\Psi}^{(3)}|3_{k}\rangle\!\!&\equiv&\!\!  
(-1)^{j+k}\bigl[ \Psi_{i}(1) \delta_{k,3-j}\delta(3,2)
\nonumber \\
& &\!\!
+ \delta_{ki}\delta(3,1)\Psi_{3-j}(2)\bigr],
\label{uPsi}
\end{eqnarray}
\begin{eqnarray}
\langle 1_{i}|\underline{\tilde{\Psi}}^{(3)}|32_{kj}\rangle\!\!&\equiv&\!\!  
(-1)^{j-1}\bigl[ \delta_{ik}\delta(1,3)\Psi_{j}(2)
\nonumber \\
& &\!\!
+ \delta_{i,3-j}\delta(1,2)\Psi_{3-k}(3)\bigr],
\label{uPsit}
\end{eqnarray}
\begin{equation}
\langle 1_{i}|\underline{\Gamma}^{(2)}|2_{j}\rangle\equiv
\Gamma^{(2)}_{ij}(1,2).
\label{Gamma^(2)}
\end{equation}
\end{subequations}
The quantity $\underline{\chi}^{(0)}$ describes independent propagation of two particles.

Using Eqs.\ (\ref{Vec3}) and (\ref{VecMat}) and noting the comments below Eq.\ (\ref{dPsi-eq}), 
we can express Eq.\ (\ref{dGPsi-eq}) as 
$
\delta \vec{G}=\frac{1}{2}\underline{\chi}^{(0)}\delta\vec{U}
-\underline{\chi}^{(0)}\underline{\Gamma}^{(4)}\delta \vec{G}
+\underline{\chi}^{(0)}\underline{\Gamma}^{(3)}\hat{\sigma}_{3}\delta \vec{\Psi}
$ 
and
$
\hat{\sigma}_{3}\delta \vec{\Psi}=\frac{1}{2}\hat{G}\underline{\tilde{\Psi}}^{(3)}\delta\vec{U}
-\hat{G}\underline{\tilde{\Gamma}}^{(3)}\delta \vec{G}
+\hat{G}\hat{\Gamma}^{(2)}\hat{\sigma}_{3}\delta \vec{\Psi}
$.
They are further transformed into
\begin{subequations}
\label{BS0}
\begin{equation}
\delta\vec{G} = \frac{1}{2}\underline{\chi}^{(4)}\delta\vec{U}+\underline{\chi}^{(4)}
\underline{\Gamma}^{(3)}\hat{\sigma}_{3}\delta\vec{\Psi},
\end{equation}
\begin{equation}
\hat{\sigma}_{3}\delta\vec{\Psi} = \frac{1}{2}\hat{\chi}^{(2)}\underline{\tilde{\Psi}}^{(3)}\delta\vec{U}-\hat{\chi}^{(2)}
\underline{\tilde{\Gamma}}^{(3)}\delta\vec{G},
\end{equation}
\end{subequations}
where $\underline{\chi}^{(4)}$ and $\underline{\chi}^{(0)}$ are defined by
\begin{subequations}
\label{chicq}
\begin{equation}
\underline{\chi}^{(4)}\equiv
\bigl(\underline{1}+\underline{\chi}^{(0)}\underline{\Gamma}^{(4)}\bigr)^{-1}\underline{\chi}^{(0)} 
=
\bigl(\underline{\chi}^{(0)-1}+\underline{\Gamma}^{(4)}\bigr)^{-1} ,
\label{chi^(4)}
\end{equation}
\begin{equation}
\hat{\chi}^{(2)}\equiv 
\bigl(\hat{1}-\hat{G}\hat{\Gamma}^{(2)}\bigr)^{-1}\hat{G} 
=\bigl(\hat{G}^{-1} -\hat{\Gamma}^{(2)}\bigr)^{-1}.
\label{chi^(2)}
\end{equation}
It is also convenient to introduce
\begin{figure}[t]
        \begin{center}
                \includegraphics[width=0.95\linewidth]{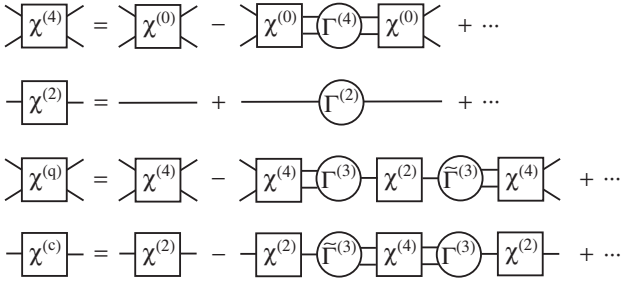}
        \end{center}
        \caption{Diagrammatic representation of Eq.\ (\ref{chicq}). Every long straight line in the second equation denotes $\hat{G}$.}
        \label{fig:Fig2}
\end{figure}
\begin{equation}
 \underline{\chi}^{({\rm q})}\equiv\left(\underline{\chi}^{(4)-1}+
\underline{\Gamma}^{(3)}\hat{\chi}^{(2)}
\underline{\tilde{\Gamma}}^{(3)}\right)^{\!\!-1},
\label{chiq}
\end{equation}
\begin{eqnarray}
 \hat{\chi}^{({\rm c})}&\equiv&\left(\hat{\chi}^{(2)-1}+
\underline{\tilde{\Gamma}}^{(3)}\underline{\chi}^{(4)}
\underline{\Gamma}^{(3)}\right)^{\!\!-1}
\nonumber \\
&=& \left(\hat{G}^{-1}-\hat{\Gamma}^{(2)}+
\underline{\tilde{\Gamma}}^{(3)}\underline{\chi}^{(4)}
\underline{\Gamma}^{(3)}\right)^{\!\!-1},
\label{chic}
\end{eqnarray}
\end{subequations}
where the superscripts $^{\rm q}$ and $^{\rm c}$ denote ``quasiparticle'' and ``condensate,'' respectively.
Figure 2 expresses Eqs.\ (\ref{chi^(4)})-(\ref{chic})  diagrammatically.
Now, we can write down the solution to Eq.\ (\ref{BS0}) in terms of $\underline{\chi}^{({\rm q})}$ and $\hat{\chi}^{({\rm c})}$ as
\begin{subequations}
\label{BS1}
\begin{equation}
\delta\vec{G} 
= \frac{1}{2}\underline{\chi}^{({\rm q})}
\bigl(\underline{1}
+\underline{\Gamma}^{(3)}\hat{\chi}^{(2)}\underline{\tilde{\Psi}}^{(3)}\bigr)\delta\vec{U},
\end{equation}
\begin{equation}
\delta\vec{\Psi} 
= \frac{1}{2}\hat{\sigma}_{3}\hat{\chi}^{({\rm c})}
\bigl(\underline{\tilde{\Psi}}^{(3)}-\underline{\tilde{\Gamma}}^{(3)}\underline{\chi}^{(4)}\bigr)\delta\vec{U}.
\end{equation}
\end{subequations}
Let us substitute Eq.\ (\ref{BS1}) into Eq.\ (\ref{calK-deriv2}) and make use of Eq.\ (\ref{VecMat})
as well as $\underline{\chi}^{({\rm q})}\underline{\Gamma}^{(3)}\hat{\chi}^{(2)}=
\underline{\chi}^{(4)}\underline{\Gamma}^{(3)}\hat{\chi}^{({\rm c})}$ in Eq.\ (\ref{chicq}).
We thereby obtain $\underline{\cal K}$ defined by Eq.\ (\ref{calK-mat}) as
\begin{eqnarray}
\underline{\cal K}&=& \underline{\chi}^{({\rm q})}
+\underline{\chi}^{(4)}\underline{\Gamma}^{(3)}\hat{\chi}^{({\rm c})}\underline{\tilde{\Psi}}^{(3)}
+\underline{\Psi}^{(3)}\hat{\chi}^{({\rm c})}
\underline{\tilde{\Gamma}}^{(3)}\underline{\chi}^{(4)}
\nonumber \\
& &
-\underline{\Psi}^{(3)}\hat{\chi}^{({\rm c})}\underline{\tilde{\Psi}}^{(3)} .
\label{ucalK}
\end{eqnarray}
This expression clearly tells us that collective modes are determined as poles of $\underline{\chi}^{({\rm q})}$ and $\hat{\chi}^{({\rm c})}$.
Note in this context that poles of $\underline{\chi}^{(4)}$ in Eq.\ (\ref{ucalK}) are cancelled 
by those of $\underline{\chi}^{(4)}$ in the denominator of $\hat{\chi}^{({\rm c})}$, as seen from Eq.\ (\ref{chic}).
It also follows from Eq.\ (\ref{chic}) that the poles of $\hat{\chi}^{({\rm c})}$ are generally not identical to those of the single-particle Green's function $\hat{G}$ due to the additional contribution $\hat{\Gamma}^{(2)}-\underline{\tilde{\Gamma}}^{(3)}\underline{\chi}^{(4)}\underline{\Gamma}^{(3)}$, in contradiction to the statement by Gavoret and Nozi\`eres.\cite{GN64} This point will be discussed in more detail below.

Equation (\ref{ucalK}) with Eqs.\ (\ref{Gamma}), (\ref{VecMat}), and (\ref{chicq}) is the main result of the present paper. The expression is formally exact, clarifies the structure of the two-particle Green's function $\underline{\cal K}$ in terms of unambiguously defined vertices, and enables us to carry out practical calculations of  $\underline{\cal K}$ for a given approximate $\Phi$ on the same footing as thermodynamic and single-particle properties. \cite{Kita09} The last point may be regarded as a definite advantage of the present formalism over the dielectric one.  \cite{SK74,WG74,Griffin93} 

Equation (\ref{ucalK}) in the coordinate representation can be used to investigate two-particle correlations of general inhomogeneous systems, including homogeneous ones.
For the latter cases, however, it is far more convenient to adopt the ``energy''-momentum representation. To be specific, vertices (\ref{Gamma}) in those cases can be expanded as
\begin{subequations}
\label{Gamma-Fourier}
\begin{eqnarray}
{\Gamma}^{(4)}_{ij,kl}(12,34)&=&\sum_{\vec{p}\vec{p}^{\,\prime}\vec{q}}
{\Gamma}^{(4)}_{ij,kl}(\vec{p},\vec{p}^{\,\prime},\vec{q})\,{\rm e}^{i(\vec{p}+\vec{q})\cdot\vec{r}_{1}-i\vec{p}\cdot\vec{r}_{2}}
\nonumber \\ & & \times {\rm e}^{i\vec{p}^{\,\prime}\cdot\vec{r}_{3}-i(\vec{p}^{\,\prime}+\vec{q})\cdot\vec{r}_{4}},
\end{eqnarray}
\begin{equation}
{\Gamma}^{(3)}_{ij,k}(12,3)=\sum_{\vec{p}\vec{q}}
{\Gamma}^{(3)}_{ij,k}(\vec{p},\vec{q})\,{\rm e}^{i(\vec{p}+\vec{q})\cdot\vec{r}_{1}-i\vec{p}\cdot\vec{r}_{2}-i\vec{q}\cdot\vec{r}_{3}},
\end{equation}
\begin{equation}
\tilde{\Gamma}^{(3)}_{i,jk}(1,23)=\sum_{\vec{p}^{\,\prime}\vec{q}}
\tilde{\Gamma}^{(3)}_{i,jk}(\vec{p},\vec{q})\,{\rm e}^{i\vec{q}\cdot\vec{r}_{1}+i\vec{p}\cdot\vec{r}_{2}-i(\vec{p}+\vec{q})\cdot\vec{r}_{3}},
\end{equation}
\begin{equation}
\Gamma^{(2)}_{ij}(1,2)=\sum_{\vec{q}}
{\Gamma}^{(2)}_{ij}(\vec{q})\,{\rm e}^{i\vec{q}\cdot(\vec{r}_{1}-\vec{r}_{2})},
\end{equation}
\end{subequations}
where $\vec{r}_{1}\equiv ({\bf r}_{1},-\tau_{1})$, $\vec{p}\equiv ({\bf p},\varepsilon_{n})$ with $\varepsilon_{n}\equiv 2n\pi T$ ($n=0,\pm 1,\cdots$),
and the summation over $\vec{p}$ denotes $T\sum_{n}\int d^{3}p/(2\pi)^{3}$. Other quantities in Eq.\ (\ref{VecMat}) can be expanded similarly. The Fourier coefficients of Eqs.\ (\ref{chi^(0)}), (\ref{u1}), (\ref{uPsi}), and (\ref{uPsit}) are thereby obtained as
\begin{subequations}
\begin{eqnarray}
&&\hspace{-10mm}
\chi^{(0)}_{ij,kl}(\vec{p},\vec{p}^{\,\prime},\vec{q})
=\delta_{\vec{p}^{\,\prime}\vec{p}}\,G_{il}(\vec{p}+\vec{q})G_{kj}(\vec{p})
 +(-1)^{k+l-1}
\nonumber \\
&&\hspace{15mm} 
\times \delta_{\vec{p}^{\,\prime},-\vec{p}-\vec{q}}\,G_{i,3-k}(\vec{p}+\vec{q})G_{3-l,j}(\vec{p}),
\end{eqnarray}
\begin{equation}
1_{ij,kl}(\vec{p},\vec{p}^{\,\prime},\vec{q})
=\delta_{il}\delta_{kj}\delta_{\vec{p}^{\,\prime}\vec{p}},
\end{equation}
\begin{equation}
\Psi^{(3)}_{ij,k}(\vec{p},\vec{q})
=(-1)^{j+k}\sqrt{n_{0}}\bigl(\delta_{k,3-j}\delta_{\vec{p},-\vec{q}}+\delta_{ki}\delta_{\vec{p},\vec{0}}\bigr),\end{equation}
\begin{equation}
\tilde{\Psi}^{(3)}_{i,jk}(\vec{p},\vec{q})
=(-1)^{j-1}\sqrt{n_{0}}\bigl(\delta_{ik}\delta_{\vec{p},\vec{0}}+\delta_{i,3-j}\delta_{\vec{p},-\vec{q}}\bigr),
\end{equation}
\end{subequations}
respectively, where $\delta_{\vec{p}^{\,\prime}\vec{p}}\equiv (2\pi)^{3}T^{-1}\delta({\bf p}'-{\bf p})\delta_{n'n}$ and $n_{0}$ denotes the condensate density. It then follows that Eqs.\ (\ref{chicq}) and (\ref{ucalK}) hold as they are in terms of the Fourier coefficients. For example, Eq.\ (\ref{chiq}) can be written explicitly as an integral equation for $\chi^{({\rm q})}_{ij,kl}(\vec{p},\vec{p}^{\,\prime},\vec{q})$ as
\begin{eqnarray}
&&\hspace{-4mm}
\underline{\chi}^{({\rm q})}(\vec{p},\vec{p}^{\,\prime},\vec{q})=\underline{\chi}^{(4)}(\vec{p},\vec{p}^{\,\prime},\vec{q})-\sum_{\vec{p}_{1}\vec{p}_{2}}\underline{\chi}^{(4)}(\vec{p},\vec{p}_{1},\vec{q})
\underline{\Gamma}^{(3)}(\vec{p}_{1},\vec{q})
\nonumber \\
&& \hspace{19mm}
\times \hat{\chi}^{(2)}(\vec{q})
\underline{\tilde{\Gamma}}^{(3)}(\vec{p}_{2},\vec{q})
\underline{\chi}^{({\rm q})}(\vec{p}_{2},\vec{p}^{\,\prime},\vec{q}),
\end{eqnarray}
where $\underline{\chi}^{({\rm q})}$, etc., are now matrices only in terms of the Nambu indices $i,j,\cdots$, which may be defined  explicitly as Eq.\ (\ref{VecMat}) without space-time arguments. 

We now compare Eqs.\ (\ref{chicq}) and (\ref{ucalK}) with the results for the two-particle Green's function $\underline{\cal K}$ by Gavoret and Nozi\`eres. \cite{GN64} 
Apparently, they found the same structure for $\underline{\cal K}$ as Eq.\ (\ref{ucalK}) above. They subsequently identified the quantity corresponding to $\hat{\Sigma}+\hat{\Gamma}^{(2)}-\underline{\tilde{\Gamma}}^{(3)}\underline{\chi}^{(4)}\underline{\Gamma}^{(3)}$ in $\hat{\chi}^{({\rm c})}$ of Eq.\ (\ref{chic}) with the single-particle self-energy as Eq.\ (3.4) of their paper, where $\tilde{M}$ and $JGGP$ on the right-hand side correspond to $\hat{\Sigma}+\hat{\Gamma}^{(2)}$ and $-\underline{\tilde{\Gamma}}^{(3)}\underline{\chi}^{(4)}\underline{\Gamma}^{(3)}$, respectively.
However, they did not provide detailed reasoning to the crucial statement. In this context, we would like to point out that their analysis of $\underline{\cal K}$ was carried out separately from that of $\hat{G}$ by only investigating its diagrammatic structure in the simple perturbation expansion, where $\hat{\Gamma}^{(2)}$, for example, may be mistaken easily for a part of the single-particle self-energy, as seen from the second diagram of Fig.\ 2.
This may be the reason why they concluded erroneously that the quantity corresponding to $\hat{\Sigma}+\hat{\Gamma}^{(2)}-\underline{\tilde{\Gamma}}^{(3)}\underline{\chi}^{(4)}\underline{\Gamma}^{(3)}$ is the single-particle self-energy. In contrast, our investigation of $\underline{\cal K}$ has been performed on the basis of
 Eq.\ (\ref{Dyson1}) for $\hat{G}$ and $\vec{\Psi}$, where the self-energy $\hat{\Sigma}$ is defined unambiguously at the single-particle level. It is thereby shown that the term $\hat{\Gamma}^{(2)}-\underline{\tilde{\Gamma}}^{(3)}\underline{\chi}^{(4)}\underline{\Gamma}^{(3)}$ should be regarded as additional contribution distinct from the single-particle self-energy.

Thus, of fundamental importance will be to clarify  how the extra ``self-energy'' $\hat{\Gamma}^{(2)}-\underline{\tilde{\Gamma}}^{(3)}\underline{\chi}^{(4)}\underline{\Gamma}^{(3)}$ in  $\hat{\chi}^{({\rm c})}$ shifts its poles from those of $\hat{G}$. In the weak-coupling limit, we can show $\Gamma^{(2)}_{ii}=0$ and $\Gamma^{(2)}_{12}(1,2) =-\Gamma^{(2)*}_{21}(1,2)=-2V({\bf r}_{1}-{\bf r}_{2})\Psi({\bf r}_{1})\Psi({\bf r}_{2})$ by
using Eqs.\ (25) and (26) of Ref.\ \onlinecite{Kita09} and Eq.\ (\ref{Gamma}) above.
Combined with
$\Sigma_{12}(1,2) =V({\bf r}_{1}-{\bf r}_{2})[\Psi({\bf r}_{1})\Psi({\bf r}_{2})-\langle\phi(1)\phi(2)\rangle]$
from the lowest-order gapless $\Phi$-derivable approximation, \cite{Kita09}
we thereby obtain
$\Sigma_{12}(1,2)+\Gamma_{12}(1,2) =V({\bf r}_{1}-{\bf r}_{2})[-\Psi({\bf r}_{1})\Psi({\bf r}_{2})-\langle\phi(1)\phi(2)\rangle]$. 
Thus, at the mean-field level, $\hat{\Gamma}^{(2)}$ merely changes the sign of the condensate (i.e., dominant) contribution to the off-diagonal self-energy.
It hence follows that, to the leading-order in the interaction,
poles of $(\hat{G}^{-1}-\hat{\Gamma}^{(2)})^{-1}$ are the same as those of $\hat{G}$.
However, they are not exactly identical due to the presence of  $\langle\phi(1)\phi(2)\rangle$.
Beyond the weak-coupling regime where the polarization contribution $-\underline{\tilde{\Gamma}}^{(3)}\underline{\chi}^{(4)}\underline{\Gamma}^{(3)}$ also becomes relevant in $\hat{\chi}^{({\rm c})}$, therefore, it is reasonable to
expect that poles of $\hat{\chi}^{({\rm c})}$ and $\hat{G}$ are generally different.
Further investigation needs to be carried out on the similarity or difference between the single-particle and collective excitations.

\end{document}